\documentclass[pra, aps, twocolumn, groupedaddress, superscriptaddress]{revtex4}
\usepackage{epsfig}
\usepackage{slashed}
\usepackage{graphicx}
\usepackage[usenames, dvipsnames]{color} 
\usepackage{psfrag}
\usepackage{graphics}
\usepackage{hyperref}
\usepackage{bm}

\begin{document}

\title{Polytropic equilibrium and normal modes in cold atomic traps}

\author{H. Ter\c{c}as}
\email{htercas@gmail.com}
\affiliation{Institut Pascal, PHOTON-N2, Clermont Universit\'e, Blaise Pascal University, CNRS, 24 Avenue des Landais, 63177 Aubi\`ere Cedex, France}
\author{J. T. Mendon\c{c}a}
\email{titomend@ist.utl.pt}
\affiliation{IPFN, Instituto Superior T\'{e}cnico, Av. Rovisco Pais 1, 1049-001 Lisboa, Portugal}

\begin{abstract}
The compressibility limit of a cold gas confined in a magneto-optical trap due to multiple scattering of light is a long-standing problem. This scattering mechanism induces long-range interactions in the system, which is responsible for the occurrence of plasma-like phenomena. In the present paper, we investigate the importance of the long-range character of the mediated atom-atom interaction in  the equilibrium and dynamical features of a magneto-optical trap. Making use of a hydrodynamical formulation, we derive a generalized Lane-Emden equation modeling the polytropic equilibirum of a magneto-optical trap, allowing us to describe the cross-over between the two limiting cases: temperature dominated and multiple-scattering dominated traps. The normal collective modes of the system are also computed.    
\end{abstract}

\maketitle

{\it Introduction}. Since the first realizations of cold atomic gases \cite{chu}, both theoretical and experimental investigations reveal that magneto-optical traps (MOT) pave a stage for very exciting and complex physical phenomena \cite{walker}. The interest in studying the basic properties of MOTs have, however, considerably decreased after the production of Bose-Einstein condensates \cite{legget,dalfovo}, as they started being used mainly as a riding horse to achieve quantum degeneracy. However, the study of the dynamical properties of MOTs have received much attention recently, which revives the investigation of the basic properties of laser cooled gases. Examples of such a growing interest can be found in the work realized by Kim {\it et al.} \cite{kim}, where a parametric instability is excited by an intensity modulated laser beam, and in the works of di Steffano and co-workers \cite{david, stefano, hennequin}, where the feedback of retroreflected laser beams can induce stochastic or deterministic chaos for a large optical thickness of the MOT.\par
A route for the most intriguing complex behavior in magneto-optical traps relies exactly on the multiple scattering of light, a mechanism which have been described since the early stages of MOTs as the principal limitation for the compressibility of the cloud \cite{dalibard, sesko}. Under these circumstances, the atoms experience a mediated long-range interaction potential similar to a Coulomb potential ($\sim 1/r$) \cite{pruvost} and the system can therefore be regarded as a one-component trapped plasma. In a series of previous works, we have put in evidence the important consequences of such plasma description of a cold atomic traps \cite{mendonca1}, whereas the formal analogy and the application of plasma physics techniques reveal to be important in the description of driven mechanical instabilities \cite{hugo} or even more exciting instability phenomena, like photon bubbles \cite{mendonca2}, phonon lasing \cite{mendonca3} or even the appearance of a roton minimum in the classical regime \cite{hugoroton}.\par
In this work, we investigate the hydrodynamic equilibrium and normal modes of cold atomic traps. For that purpose, we combine the effects of multiple scattering (described by the Coulomb-like potential) and the thermal fluctuations inside the system, which can be cast in the form of a polytropic equation of state. We derive a generalized Lane-Emden equation for the equilibrium density profiles and calculate the corresponding solutions. It is shown that the long-range interactions significantly change the Maxwell-Boltzmann equilibrium of a thermal gas. We linearize the equations to calculate the normal modes of the system for both small and large clouds, respectively dominated by thermal and multiple-scattering effects. We describe the collective modes of the system in both temperature limited and multiple-scattering regimes of the gas. \par
{\it Polytropic hydrodynamics and the generalized Lane-Emde equation}. A simple description of a cold gas in the presence of long-range interactions can be done using a set of hydrodynamic equations considered in our previous paper \cite{mendonca1}
\begin{equation}
\frac{\partial n}{\partial t}+\bm{\nabla}\cdot (n\mathbf{v})=0
\label{eq:cont}
\end{equation} 
\begin{equation}
\frac{\partial\mathbf{v}}{\partial t}+(\mathbf{v}\cdot\bm{\nabla})\mathbf{v}=-\frac{\bm{\nabla}P}{Mn}+\frac{\mathbf{F}_{t}}{M}+\frac{\mathbf{F}_{c}}{M},
\label{eq:momentum}
\end{equation}
\begin{equation}
\bm{\nabla}\cdot\mathbf{F_c}=Qn
\label{eq:poisson},
\end{equation}
where $n$ and $\mathbf{v}$ represent the gas density and velocity, respectively, and $M$ is the atomic mass.  Here, $\mathbf{F}_{c}$ is the collective force and $Q=(\sigma_{R}-\sigma_{L})\sigma_{L}I_{0}/c$ represents the square of the effective {\it electric charge} of the atoms \cite{mendonca1, pruvost}, with $c$  being the speed of light, $I_{0}$ the total intensity of the six laser beams. $\sigma_{R}$ and $\sigma_{L}$ represent the emission and absorption cross sections respectively \cite{walker}. The term $\mathbf{F}_{t}=-\bm{\nabla}U$ stands fot the trapping force. The trapping potential is assumed to be harmonic
\begin{equation}
U(\mathbf{r})= \frac{1}{2}\kappa r^2,  
\label{single}
\end{equation}
and $\kappa$ represents the spring constant (in the low saturation Doppler limit, the spring constant is approximately given by $\kappa=\alpha \mu_{B}\nabla B/\hbar k$ - for the sake generality, we shall consider a generic value of $\kappa$ in the remainder of the paper). Here, $\mu_{B}$ represents the Bohr magneton, $\kappa=\kappa(\delta, I_{0}/I_{s})$ is the spring constant, $\alpha=\alpha(\delta, I_{0}/I_{s})$ is the friction coefficient, $\delta$ is the laser detuning, $I_{s}$ is the atomic saturation intensity and $\nabla B=\vert \bm{\nabla}B\vert$ represents the magnetic field gradient \cite{mendonca1}. The harmonic trapping potential \ref{single} is valid for a MOT of maximum radius $R_\textrm{max}=\vert \delta\vert/(\gamma_M\nabla B)$, where $\gamma_M=g_J\mu_B/\hbar$ and $g_J$ is the LandŽ factor. Under typical experimental conditions, $\nabla B\simeq 10$ G cm$^{-1}$, $\delta=-2\Gamma_0$ (with $\Gamma_0$ standing for the atomic transition life-time), we find $R_\textrm{max}~\sim 0.5$ cm. Above that limit, self-sustained mechanical instabilities take place \cite{labeyrie}, so we should avoid this situation in the present work. \par
In the absence of a microscopic theory of the ultra-cold gas, we assume a polytropic equation of state for the MOT, of the form
\begin{equation}
P=C_{\gamma}(T) n^\gamma,
\label{eq:polytropic}
\end{equation}
where $\gamma$ is the polytropic exponent and $C_{\gamma}(T)$ is a certain function of the temperature $T$. The hydrodynamic equilibrium condition applied to Eqs. (\ref{eq:cont}, \ref{eq:momentum}, \ref{eq:poisson}) simply yields
\begin{equation}
\bm{\nabla}\cdot\frac{\bm{\nabla} P}{n}= 3M\omega_{0}^2-Qn,
\label{eq:equilibrium1}
\end{equation}
where $\omega_{0}=\sqrt{\kappa/M}$ is the trapping frequency. Assuming radial symmetry, the density is given by $n=n(0) \theta(r)$, where $n(0)$ represents the peak density. Putting Eqs. (\ref{eq:polytropic}) and (\ref{eq:equilibrium1}) together, one easily obtains
\begin{equation}
\gamma \frac{1}{\xi^2}\frac{d}{d\xi}\left(\xi^2\theta^{\gamma-2}\frac{d\theta}{d\xi}\right) - \Gamma \theta +1=0,
\label{eq:equilibrium2}
\end{equation}
where $\Gamma=Q n(0)/3M\omega_{0}^2C_{\gamma}(T)$ represents the ratio of interaction to kinetic energy (coupling parameter). The distance $\xi=r/a_{\gamma}$ is given in units of a generalized Wigner-Seitz radius
\begin{equation} 
a_{\gamma}=\sqrt{\frac{3M\omega_{0}^2}{C_{\gamma}(T)}}n(0)^{-(\gamma-1)/2}.
\end{equation}
Equation (\ref{eq:equilibrium2}) corresponds to a generalization to the Lane-Emden equation derived to study astrophysical fluids \cite{harko}. The important modifications in our model include both the trapping and the long-range interaction induced by multiple scattering. The present model allows us to generalize the theory of work by Walker et al. \cite{walker} by relating $\gamma$ to experimentally accessible density profiles. The case of rotating clamps of atoms, however, is not included in this paper and will be considered in a separated publication \cite{mendnovo}. In what follows, we examine the analytical solutions of Eq. (\ref{eq:equilibrium2}) for some limiting cases, and compare them with numerical solutions. \par

\begin{figure}[t!]
\includegraphics[scale=0.65]{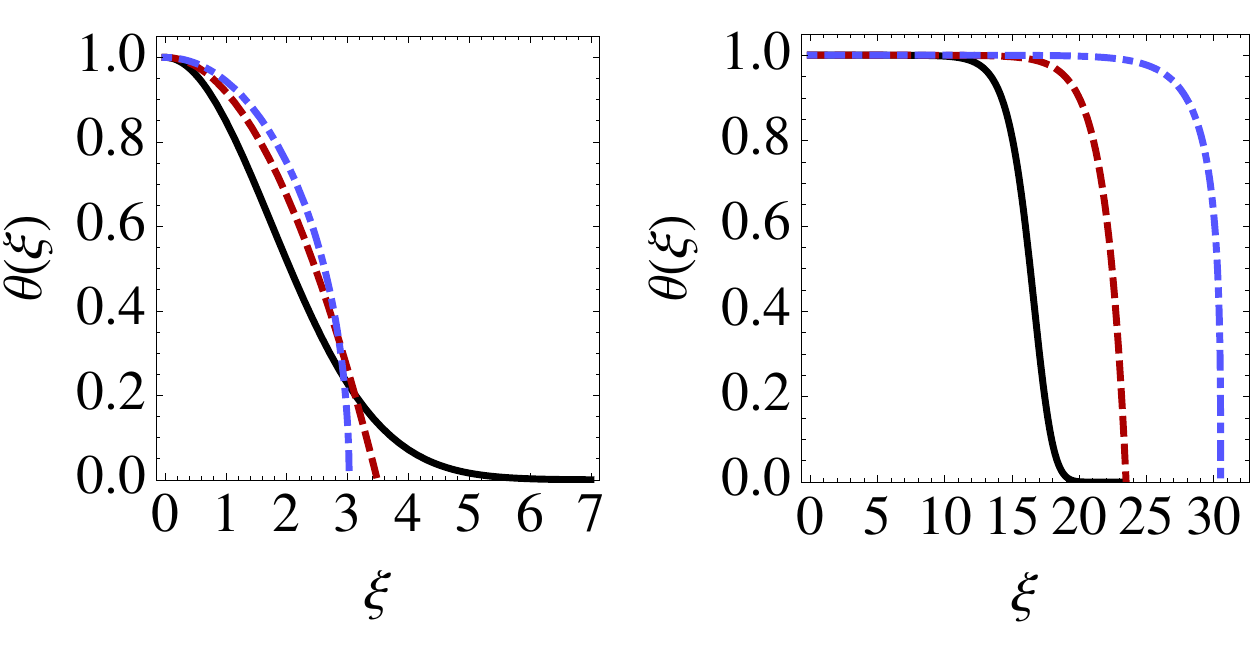}
\caption{Effect of the long-range interaction on the density profile for different polytropic exponents. The left panel depicts the case where thermal effects dominates ($\Gamma=0$), while the right panel illustrates the case where multiple scattering dominates ($\Gamma=0.99$). The black (thick) line depicts the normalized density profile for the isothermal case, $\gamma=1$. Red (dashed) and blue (dot-dashed) lines are obtained for $\gamma=2$ and $\gamma=4$, respectively.}
\label{equilibrium}
\end{figure}

{\it Multiple scattering regime: cold plasma equilibrium.} In the multiple scattering regime, typically achieved for a number of particles above $N\sim 10^4-10^5$, the MOT is essentially dominated by the collective forces \cite{walker}. In that case, $\Gamma \rightarrow 1$ and one can safely neglect the effects of the pressure. By setting $\gamma=0$ (and consequently $C_{0}(T)=1$), Eq. (\ref{eq:equilibrium2}) simply yields the so called \textit{water-bag} equilibrium profile
\begin{equation}
\theta(\xi)=\Gamma^{-1}\Theta(\xi-\xi_1),
\label{waterbag}
\end{equation}
where $\xi_1=3/(4\pi\Gamma)^{1/3}$ is the Lane-Emden radius of the MOT. In physical units, it corresponds to a MOT of radius
\begin{equation}
R=\left(\frac{3 N}{4\pi n_0}\right)^{1/3}\approx 0.62 N^{1/3}a_0.
\end{equation}
This exactly corresponds to the scaling law observed in the experiments of Ref. \cite{gattobigio}. For typical $^{85}$Rb MOT, $\delta = -\Gamma_0/2$ and $\nabla B = 10$  Gcm$^{-1}$, we get a density $n_0 = 3 \times 10^9$ cm$^{-3}$. This density would be reached for $N = 1500$ atoms already. This means that in typical MOTs, the large density limit very easily holds. In this case, the restoring frequency and the effective plasma frequency $\omega_P=\sqrt{Q n_0 / M}$  are related as
\begin{equation}
\omega_0=\frac{\omega_p}{\sqrt{3}}.
\end{equation}
We will later see that this is related with the Mie mode, which is a natural consequence of treating the system as trapped one-component plasma. In Fig. \ref{equilibrium}, it is represented the numerical solution to Eq. (\ref{eq:equilibrium2}) for different polytropes. We observe that the Gaussian profile is modified when $\Gamma$ is increased towards a water-bag profile. 
For traps containing a small number of particles ($N\lesssim 10^4$), the effects of the multiple scattering can be neglected. In that case, we set $\Gamma \rightarrow 0$ and obtain the following polytropic equilibrium
\begin{equation}
\theta(\xi)=\left(1-\frac{\gamma-1}{6\gamma}\xi^2 \right)^{1/(\gamma-1)}.
\label{eq:thermal}
\end{equation}
For the interesting case of an isothermal gas, $\gamma=1$ and $C_{1}=k_{B}T$, this simply leads to the Maxwell-Boltzman equilibrium
\begin{equation}
n_{1}(r)=n(0)e^{-V_{t}/k_{B}T}=n(0)e^{-r^2/6a_{1}^2},
\end{equation}
where $a_{1}=\sqrt{3m\omega_{0}^2/k_{B}T}$.
\par
{\it Normal modes}.
We now discuss the case of the localised oscillations, or \textit{normal modes}, in trapped gases. By linearising the set of fluid equations (\ref{eq:cont}), (\ref{eq:momentum}) and (\ref{eq:poisson}) readily yield
\begin{equation}
\begin{array}{c}
\displaystyle{-\omega^2\delta n-\frac{\gamma C(T)}{m}\bm{\nabla}\cdot \left(n_0^{\gamma-1}\bm\nabla \delta n\right)=\frac{1}{m}\bm{\nabla}\cdot\left[n_{0}(r) \bm\nabla{\delta \phi}_{c}\right], \label{eq_modes1}}\\\\
\displaystyle{\nabla^2 \delta\phi_{c}=-Q\delta n,}
\end{array}
\label{eq_modes2}
\end{equation} 
where we have used $\bm\nabla\delta P\simeq \gamma C(T)n_0^{\gamma-1}\bm\nabla\delta n$. Here, we have assumed that the collective force can be derived from a potential, i.e., $\mathbf{F}_c=-\bm\nabla \phi_c$. We now define the auxiliary quantity $\eta$, defined as $\delta n= (1 / 4\pi r^2) d\eta / d r$, which together with Eq. (\ref{eq_modes2}) implies
\begin{equation}
\frac{d}{dr}\delta \phi_c=-\frac{Q}{4\pi r^2}\eta.
\end{equation}
Making proper substitutions, the linearized equations can be put together and result in a single expression
\begin{equation}
\begin{array}{c}
\displaystyle{-\frac{\gamma C(T)n(0)^{\gamma-1}}{m}\frac{1}{r^2}\frac{d}{d r}\left[r^2n_0(r)^{\gamma-1}\frac{d}{dr}\left(\frac{1}{r^2}\frac{d\eta}{d r} \right)\right]}\\
\displaystyle{-\omega^2\frac{1}{r^2}\frac{d\eta}{d r}+\omega_p^2\frac{1}{r^2}\frac{d}{dr} \left[n_0(r) \eta\right]=0,}
\end{array}
\end{equation}
where we have used $\omega_p^2=Qn(0)/m$. In a dimensionless form, one obtains
\begin{equation}
\begin{array}{c}
\displaystyle{-\omega^2\frac{1}{\xi^2}\frac{d\delta \theta}{d \xi}-\frac{1}{2}\gamma \omega_0^2\frac{1}{\xi^2}\frac{d}{d\xi} \left[\xi^2\theta\frac{d}{d\xi}\left(\frac{1}{\xi^2}\frac{d\delta \theta}{d \xi} \right)\right]}\\
\displaystyle{+\omega_p^2 \frac{1}{\xi^2}\frac{d}{d\xi}\left(\theta^{1/(\gamma-1)} \delta\theta\right)=0.}
\end{array}
\label{eq:modestotal}
\end{equation}
Solutions to the latter eigenvalue problem depend on the details of the equilibrium $\theta(\xi)$ considered. The general case involves a numerical solution, for which we hereby provide analytical solutions for some limiting cases. For small clouds, lying in the temperature limited regime ($N<10^4$), the effects of multiple scattering may be neglected and, therefore, one can set $\omega_p=0$ in the Eq. (\ref{eq:modestotal}). Using the equilibrium profile in Eq. (\ref{eq:thermal}), the eigenvalue problem yields
\begin{equation}
-\omega^2\delta n-\frac{1}{2}\gamma \omega_0^2\frac{1}{\zeta^2}\frac{d}{d\zeta} \left[\left(1-\zeta^2\right)\zeta^2\delta n \right]=0
\label{thermal_modes},
\end{equation}
where we have performed a change of variables, $\zeta=\sqrt{6\gamma/(\gamma-1)}\xi$. The solution can be given in terms of the ansatz \cite{footnote} 
\begin{equation}
\delta n=\sum_{n,\ell}a_{n\ell} \zeta^{2n+\ell}.
\end{equation}
Replacing this in Eq. (\ref{thermal_modes}), one easily obtains a recurrence relation for the coefficients $a_{n,\ell}$ which is found to converge provided that
\begin{equation}
\omega^2=\omega_0^2\left\{\ell+2n\left[(\gamma-1)\left(n+\ell+1/2\right)+1\right]\right\}.
\end{equation}
This result resembles the solution given by Stringari for the oscillations of a Bose-Einstein condensate ($\gamma=2$) in a spherical harmonic trap \cite{stringari1}. For comparison, we remind the result known for the free BEC in the collisionless regime ($\gamma=1$), $\omega=\omega_0(2n+\ell)$. The case $\gamma<1$ is naturally unstable. Pure surface modes ($n=0$), which may eventually be easier detectable experimentally, have the following frequencies 
\begin{equation}
\omega_S=\omega_0\sqrt{\ell}.
\end{equation}
On the other hand, breathing modes ($\ell=0$) are also theoretically possible in small traps, with frequencies given by
\begin{equation}
\omega_B=\omega_0\sqrt{2n+(\gamma-1)(n+1/2)}
\end{equation}
Because experimental techniques usually make possible the measurement of frequencies in a very precise manner, these results can be very useful, as they relate the polytropic exponent $\gamma$ with the mode frequencies. \par

In the deep multiple scattering regime, we can model the equilibrium by a water-bag solution in Eq. (\ref{waterbag}). In that case, one readily obtains
\begin{equation}
(3\omega_{0}^2-\omega^2)\delta n=0,
\label{eq:breathing}
\end{equation} 
which corresponds to a \textit{breathing mode} in a system with long-range interactions
\begin{equation}
\omega_{B}=\omega_p=\sqrt{3}\omega_{0}.
\end{equation}
The latter result is very well-known in plasma physics and corresponds to an uncompressional monopole oscillation of the system at the classical plasma frequency $\omega_{p}$. However, this solution is not unique. Manipulation of Eqs. (\ref{eq:cont}), (\ref{eq:momentum}), and (\ref{eq:poisson}) also yields
\begin{equation}
\bm{\bm{\nabla}} \cdot \left[\epsilon(\omega)\bm{\bm{\nabla}}{\delta \phi}_{c}^{in}\right]=0 \quad \mbox{with}\quad  \epsilon(\omega)=1-3\frac{\omega_{0}^2}{\omega^2},
\label{eq:modes3}
\end{equation} 
which holds in the interior region of the MOT, $r<R$. Outside the MOT ($r>R$), the collective force should not vary, so we have $\bm{\nabla}^2\delta \phi_{c}^{\textrm{out}}=0$. The general solution to Eq. (\ref{eq:modes3}) is therefore given by
\begin{equation}
\begin{array}{cc}
\displaystyle{{\delta \phi}_{c}^{{\rm in}}(r)=\sum _{\ell,m}a_{\ell,m}~r^\ell Y_{\ell}^m(\theta,\varphi)}\\
\displaystyle{{\delta \phi}_{c}^{{\rm out}}(r)=\sum_{\ell,m} b_{\ell,m}~r^{-(\ell+1)} Y_{\ell}^m(\theta,\varphi)}.
\end{array}
\end{equation}
Imposing regular continuity conditions at the surface $r=R$, 
\begin{eqnarray}
\phi_{c}^{{\rm in}}(R)&=&\phi_{c}^{{\rm out}}(R)\nonumber\\
\left. \frac{d}{d r}\phi_{c}^{{\rm  in}}(r)\right \vert_{r=R}&=&\left. \frac{d}{d r}\phi_{c}^{{\rm out}}(r)\right\vert_{r=R},
\label{boundary_cond}
\end{eqnarray}
one obtains the frequencies corresponding to incompressible \textit{surface} modes
\begin{equation}
\omega_{S}=\omega_{0}\sqrt{\frac{3\ell}{2\ell+1}}.
\label{eq:surface}
\end{equation}
A similar result can be obtained from the Mie theory for scattering in the context of surface plasmon-polaritons (check Ref. \cite{pitarke} for an understandable review). We remark here that the frequency of surface modes are bounded between the Mie and the surface plasmon resonances,  $\omega_p/\sqrt{3}<\omega<\omega_p/\sqrt{2}$, totally differing from the Tonks-Dattner resonances found described in our previous work \cite{mendonca1}. We expect that this feature can be very easily observed in fluorescence measurements, for which a plateau in a frequency resolved measurement would appear. Notice that the modes in Eq. (\ref{eq:surface}) do not depend on the size of the cloud $R$. This is a result of neglecting the dispersion, introduced by thermal fluctuations. \par
{\it Retarded surface modes in large traps}.
\begin{figure}[t!]
\centering
\includegraphics[scale=0.8]{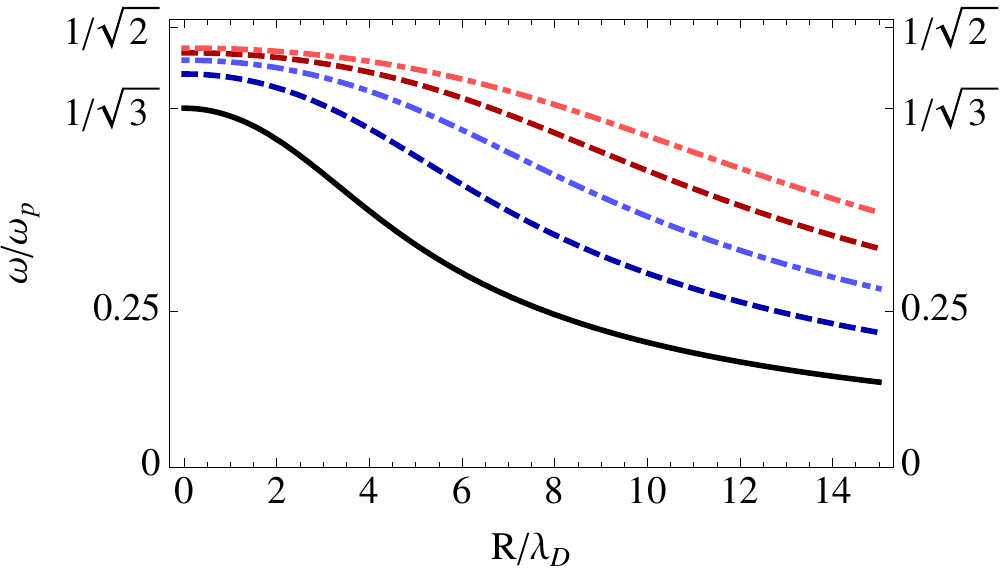}
\caption[Retardation effects on the surface modes]{Effect of the retardation on the surface modes.  From bottom to top: $\ell=1,2,3,4 \mbox{ and }5$.}
\label{fig_modes}
\end{figure}
The previous calculation of the frequencies of surface oscillations took place in the cold fluid limit. In reality, the thermodynamic pressure is always present and therefore  the effects of dispersion must be considered. A straightforward manipulation of the fluid equations yields an equation for the collective potential as follows
\begin{equation}
\left(\nabla^2+k^2_{\rm in }\right)\delta\phi_c^{\rm in}=0,\quad \nabla^2\delta\phi_c^{\rm out}=0,
\label{eq:modes2}
\end{equation}
where $k_{\rm in}=\sqrt{\epsilon(\omega)}\omega/u_s$ and $\epsilon(\omega)$ is the frequency response given in (\ref{eq:modes3}). The general solution to Eq. (\ref{eq:modes2}) reads, after ruling out exponential growth at the origin,  

\begin{equation}
\begin{array}{cc}
\displaystyle{{\phi}_{c}^{\rm in}(r)=\sum _{\ell,m}a_{\ell,m}~j_\ell(k_{\rm in}r) Y_{\ell}^m(\theta,\varphi)}\\
\displaystyle{{\phi}_{c}^{\rm out}(r)=\sum_{\ell,m} b_{\ell,m}~r^{-(\ell+1)} Y_{\ell}^m(\theta,\varphi)},
\end{array}
\end{equation}
where $j_\ell(z)$ represents the spherical Bessel functions of the first kind. Imposing the boundary conditions in (\ref{boundary_cond}), one finally gets the eigenvalues through the following condition
\begin{equation}
 \frac{j_\ell(\sqrt{\epsilon(\omega)}\omega x)}{\sqrt{\epsilon(\omega)}j_\ell'(\sqrt{\epsilon(\omega)}\omega x)}= -\frac{\ell+1}{x},
\end{equation}
 where $x=R/\lambda_D$. The numerical solution to the latter is plotted in Fig. (\ref{fig_modes}). Notice that the frequencies tend to decrease their values as $R$ increases. Such a feature is often referred to as {\it retardation} in the context of surface plasmon-polaritons \cite{pitarke}. In real-life experiment, the effect of retardation can be tested by taking the spectrum of the Fourier transform density profile. By increasing the number of atoms $N$ in the cloud, a broadening of the plateau $\omega_p/\sqrt{3}<\omega<\omega_p/\sqrt{2}$ should be observed, as the difference between two consecutive surface modes increase with $R$ (see Fig. \ref{fig_modes}). \par
{\it Conclusions}.
We have used a hydrodynamic polytropic description of the magneto-optical trap and derived the equilibrium condition. The equilibrium density is given as the solution of a generalized Lane-Emden equation. We have also computed the spectrum of the normal modes in the two relevant limits of the system (dominated by temperature, or dominated by multiple scattering). We have established a distinction between radial and surface modes as a function of the polytropic exponent $\gamma$ and showed that a retardation (decrease of frequency) of the surface modes is associated with the thermal effects, deviating from the cold plasma result. The importance of the presents results is twofold: first, a polytropic equation of state, which phenomenologically models a large class of hydrodynamic problems, can be easily confirmed experimentally, both by measuring the density profiles and by determining the spectrum of the normal modes; second, our problem may establish an important bridge to investigate astrophysical systems in the laboratory, specially in what concerns the stability and dynamics of the Lande-Endem equation. \par
This work was supported by Marie Curie Actions through the grant number FP7 ITN ``Spin-Optronics" (237252).


\bigskip


\begin{thebibliography}{10}
\bibitem{chu} S. Chu, Rev. Mod. Phys. {\bf 70}, 685 (1998); C. Cohen-Tannoudji, {\textit ibid.} {\bf 70}, 707 (1998); W. D. Philips, {\textit ibid.} {\bf 70}, 721 (1998). 
\bibitem{walker} T. Walker, D. Sesko, and C. Wieman, Phys. Rev. Lett. {\bf 64}, 408 (1990).
\bibitem{legget} A. J. Legget, Rev. Mod. Phys. {\bf 73}, 307 (2001).
\bibitem{dalfovo} F. Dalfovo {\textit et al.}, Rev. Mod. Phys. {\bf 71}, 463 (1999).
\bibitem{kim} K. Kim,  H.-R. Nohm and W. Jhe, {Opt. Comm.} {\bf 236} {349} (2004).
\bibitem{david} D. Wilkowski , J. Ringot , D. Hennequin , and J. C. Garreau, Phys. Rev. Lett. {\bf 85} {1839} (2000).
\bibitem{stefano} A. di Stefano A., M. Fauquembergue, P. Verkerk, and D. Hennequin , Phys. Rev. A {\bf 67}, 033404 (2003);
A. di Stefano, P. Verkerk, and D. Hennequin , {Eur. Phys. J. D} {\bf 30}, {243} (2004).
 \bibitem{hennequin} D. Hennequin ,  {Eur. Phys. J. D} {\bf 28}, {135} (2004).
  \bibitem{dalibard} J. Dalibard, Opt. Commun. {\bf 68}, 203 (1988).
\bibitem{sesko} D. W. Sesko D. W., T. G. Walker, and C. E. Wieman,  J. Opt. Soc. Am. B, {\bf 8} {946} (1991).
\bibitem{pruvost} L. Pruvost et al., {Phys. Rev. A}, {\bf 61}, 053408 (2000).
\bibitem{mendonca1} J.T. Mendon\c{c}a, R. Kaiser, H. Ter\c{c}as and J. Loureiro, {\sl Phys. Rev. A} {\bf 78}, 013408 (2008); {Phys. Rev. A}, {\bf 81}, 023421 (2010). 
\bibitem{labeyrie} G. Labeyrie, F. Michaud, and R. Kaiser, Phys. Rev. Lett. {\bf 96}, 023003 (2006).
\bibitem{hugo} H. Ter\c{c}as, J.T. Mendon\c{c}a and R. Kaiser, { Europhys. Lett.}, {\bf 89}, 53001 (2010).
\bibitem{mendonca2} J.T. Mendon\c ca and R. Kaiser, {Phys. Rev. Lett.}, {\bf 108}, 033001 (2012).
\bibitem{mendonca3}  J.T. Mendon\c ca, H. Ter\c cas, G. Brodin and M. Marklund, Eur. Phys. Lett. \textbf{91}, 33001 (2010).
\bibitem{hugoroton} H. Ter\c cas, J. T. Mendon\c ca, and V. Guerra, Phys. Rev. A {\bf 86}, 053630 (2012).
\bibitem{harko} C. G. B\"omer and T. Harko, J. C. Ast. Phys. {\bf 06}, 025 (2007).
\bibitem{gattobigio} G. L. Gattobigio, T. Pohl, G. Labeyrie, and R. Kaiser, Phys. Scr. {\bf 81}, 025301(2010).
\bibitem{mendnovo} J. T. Mendon\c ca et al, in preparation (2013).
\bibitem{footnote} More formally, the solution can be found directly by inspection, where the departure from equilibrium is found to be $\delta n\propto r^\ell (1-\zeta^2)^{1/(\gamma-1)}F[-n,n+\ell +1/(\gamma-1),\ell+3/2;\zeta^2]$, where $F[a,b,c;z]$ represents the hypergeometric function.
\bibitem{stringari1} S. Stringari, Phys. Rev. Lett. {\bf 77}, 2360 (1996).
\bibitem{pitarke} J. M. Pitarke, V. M. Silkin, E. V. Chulkov and P. M. Echenique, Rep. Prog. Phys. {\bf 70}, 1 (2007).

\end{thebibliography}
\end{document}